\documentclass[twocolumn]{aastex631}
\usepackage{textcomp}
\usepackage{amssymb}
\usepackage{rotating}

\newcommand{\fermi}{\textit{Fermi}}
\newcommand{\gr}{$\gamma$-ray}

\begin{document}

\title{Testing Millisecond Pulsars as the Source of the Galactic Center Excess Gamma-Ray emission}

\author{Yi Xing}
\affiliation{Key Laboratory for Research in Galaxies and Cosmology, Shanghai Astronomical Observatory, Chinese Academy of Sciences, 80 Nandan Road, Shanghai 200030, China; yixing@shao.ac.cn; wangzx20@ynu.edu.cn}

\author{Zhongxiang Wang}
\affiliation{Department of Astronomy, School of Physics and Astronomy,
Yunnan University, Kunming 650091, China}
\affiliation{Key Laboratory for Research in Galaxies and Cosmology, Shanghai Astronomical Observatory, Chinese Academy of Sciences, 80 Nandan Road, Shanghai 200030, China; yixing@shao.ac.cn; wangzx20@ynu.edu.cn}

\author{Feng Huang}
\affiliation{Department of Astronomy, Xiamen University, Zengcuo'an West Road, Xiamen 361005, China}

\begin{abstract}
The Galactic Center Excess (GCE) $\gamma$-ray emission detected with the Large
Area Telescope onboard the {\it Fermi Gamma-ray Space Telescope} has been
considered as a possible sign for dark matter (DM) annihilation, but other
	possibilities such as the millisecond pulsar (MSP) origin have also 
	been suggested. As a spectral
fitting method, constructed based on properties of $\gamma$-ray MSPs, has been
developed, we apply this method to the study of the GCE emission for the purpose
of probing the MSP origin for the GCE. A number of $\sim$1660 MSPs can provide
a fit to the spectrum of the GCE emission upto $\sim$10\,GeV, but the higher
energy part of the spectrum requires additional emission components. We further
carry out a stacking analysis of 30--500\,GeV data for relatively nearby
$\gamma$-ray MSPs,
and the resulting flux upper limits are still lower than those of the GCE 
	emission.
We consider the single DM annihilation channel $\tau^{+}\tau^{-}$ or 
	channel $b\bar{b}$,
or the combination of the two for comparison, and find they generally can 
provide better fits than MSPs. Combination of MSPs plus a DM channel
are also tested, and MSPs plus the DM channel $b\bar{b}$ can always 
provide better fits. Comparing this combination case to the pure DM 
channel $b\bar{b}$, the MSP contribution is found to be marginally needed.
\end{abstract}

\keywords{Galactic center (565); Gamma-rays (637); Millisecond pulsars (1062);
Dark matter (353)}

\section{Introductions}

More than 80\% of the matter in the Universe is dark matter (DM). While its 
nature is under intense investigation, one type of the most compelling 
candidates is the Weakly Interacting Massive Particles (WIMP; 
e.g., \citealt{ber+05}).  These DM particles are expected to annihilate, 
producing $\gamma$-ray--emission signals through possible channels that 
depend on the final states with different Standard Model particles 
(e.g., \citealt{fun15,arc+18}). This possibility has
motivated indirect searches for the DM in the \gr\ band. 
The Large Area Telescope (LAT), the main instrument onboard {\it
the Fermi Gamma-ray Space Telescope}, scans the whole sky every three hours 
in the energy range from 
tens of MeV to $\sim$1 TeV \citep{atw+09}. The data collected with LAT
well serve the purpose of performing indirect searches for DM-annihilation 
signals across the entire sky.

Targets for indirect searches are those regions supposed to have
high DM densities.  The Galactic Center (GC) is one of them, actually expected
to be the brightest source at $\gamma$-rays from WIMP annihilations 
(e.g., \citealt{spr+08,fun15}).  Diffuse excess \gr\ emission, possibly 
related to the annihilations around the GC, was first reported  
after one year's observations of \fermi-LAT \citep{hg11}. Since then,
the detection of the so-called Galactic Center Excess (GCE) emission has been
reported in quite a few studies (e.g., \citealt{ak12,ccw15,dfh+16,aaa+17}). 
The GCE is generally found to be spherically symmetric, 
having an extension of approximately 2 kpc from the GC, and have a spectrum 
peaking in the energy range of $\sim$1--3 GeV. These properties are consistent 
with the predictions of the DM annihilation models; more specifically, 
the DM particles of several tens of GeV could annihilate to pairs of such 
as b-quarks or $\tau$-leptons, giving rise to a \gr\ spectrum 
that is able to fit the GCE emission \citep{hg11,ak12,dfh+16,hes16,d21}.

Alternatively, there is another possibility for the origin of the GCE often
discussed.  Thousands of unresolved millisecond pulsars (MSPs; e.g., 
\citealt{bk15,bar+16,abb+18}) could exist in the GC region, considered to be
formed in dense globular clusters and deposited in the GC region due to
the cluster evaporation and gravitational tidal disruption \citep{bk15,fra+18}.
Thanks to \fermi, MSPs are known as a major class of \gr-emitting sources
in our Galaxy (e.g., \citealt{3pc23}).
The \gr\ spectral shape of the GCE is similar to those of MSPs in the low-energy
GeV band, and also several studies have suggested that the GCE morphology is 
compatible with the stellar distribution in the Galactic bulge 
(e.g., \citealt{mac+18,aba+20}), which would be similar to 
Galactic globular clusters that naturally contain many MSPs  
(e.g., \citealt{mac+18}).

We have developed a method that constructs a spectral template for the \gr\ 
emission of a number of MSPs in a given region and successfully applied
it to explaining the \gr\ emissions of the Galactic globular clusters
\citep{wu+22,zxw22,zxw+23}. In this method, characteristic properties of
\gr\ spectra, 
the distribution of spin periods, characteristic ages, and the 
relation between \gr\ emission efficiencies and characteristic ages 
of the known \gr\ MSPs are taken into account.
The method can also be applied to 
fitting large-scale \gr\ emission such as from the GCE or 
nearby galaxies. Given the possible MSP origin for the GCE, we have
carried out the study by applying our method. Since it can be easily noted
that \gr\ emission of MSPs generally have a cutoff at $\sim$1--2\,GeV energies
(e.g., \citealt{xw14,wu+22}), the GCE emission contains much higher 
energy part and thus we have also tested to include the contributions from
both the MSPs and the DM annihilations in our study.

In this paper, we report the results of the study.
Below we first describe the \fermi\ LAT data and analysis for 
the GCE emission in Section \ref{sec:data}. 
The procedures of the morphology study and the GCE spectral fitting are presented in Section~\ref{sec:ms} and Section~\ref{sec:sf}, respectively.
The results are discussed and summarized in Section~\ref{sec:dis}.

\section{\textit{FERMI} LAT Data and Analysis}
\label{sec:data}

\subsection{Data}
\label{sec:obs}

We selected the LAT events from the updated \textit{Fermi} Pass 8 database.
The region of interest (RoI) was defined as that with $|l|\leq 20^{\circ}$  
and $1^{\circ}\leq|b|\leqq20^{\circ}$, where $l$ and $b$ are the Galactic 
longitude and latitude respectively.
The events within $|b|=1^{\circ}$ were excluded to avoid the strong emission 
from the Galactic disk, which would cause large flux uncertainties on the GCE.
The time period of the LAT data was 15\,yrs
from 2008-08-04 15:43:39 (UTC) to 2023-08-22 04:06:35 (UTC).
The \textit{ULTRACLEANVETO} event class of the data was used, which is 
recommended for studies of diffuse sources.
Both the \textit{Front} and \textit{Back} event types were used.
We included the events with zenith angles $\leq$90 deg, which 
prevents the Earth's limb contamination, and excluded the events with 
quality flags of `bad'. 
These two event-selection criteria are recommended by the LAT team\footnote{\footnotesize http://fermi.gsfc.nasa.gov/ssc/data/analysis/scitools/}.

\subsection{Source Model}
\label{sec:da}

\subsubsection{Catalog Sources}

We included all sources within 30 deg centered at the Galactic Center (GC) 
in the source model. The positions and the spectral parameters of the 
sources are provided in the recently updated \textit{Fermi} LAT 14-year 
source catalog (4FGL-DR4; \citealt{4fgl-updated4}).
There were 1037 point sources and 27 extended sources.
When constructing the \fermi\ LAT source catalog, the weights 
were introduced in the maximum likelihood analysis \citep{4fgl,4fgl-updated4}. 
Events below 316 MeV contributed little with low weights 
(see Figure 22 in \citealt{4fgl}), especially in the region near the Galactic 
plane. In addition, only \textit{Front} events were used in $<$316 MeV band 
in order to avoid the poor PSF and the contamination due to the Earth's
limb \citep{4fgl}. Because of these, we only used the LAT events in 300 MeV 
to 500 GeV in the following analyses, so that the large uncertainties 
in modeling of the catalog sources and the background diffuse emission 
in the low energy band were avoided.

\subsubsection{Large-Scale Diffuse Sources}

In the \gr\ sky, the main large-scale diffuse sources include the Galactic 
and extragalactic diffuse emission, the \fermi\ bubbles, the Loop I, and 
the GCE component.  The spectral model gll\_iem\_v02\_P6\_V11\_DIFFUSE.fit 
was used for the Galactic diffuse emission, which was initially often 
adopted for the studies of diffuse \gr\ emission
(e.g., \citealt{hs13,day+16}). 
It consists of the $\pi^{0}$, the bremsstrahlung, and the inverse Compton 
scattering (ICS) components, and has only 1 free parameter, the normalization.
The spectral file iso\_P8R3\_ULTRACLEANVETO\_V3\_v1.txt was used for 
the extragalactic diffuse emission.
We adopted the flat intensity spatial templates given in \citet{su+10} 
to describe the \fermi\ bubbles and Loop I. The latter has a large size with
$b > 25$\,deg, which is outside of our RoI.
In the analysis, we found that whether or not including this component did 
not significantly affect the spectral results, and thus we removed the Loop I 
template from our source model. 

For the GCE emission, it was claimed in many studies 
(e.g., \citealt{ccw15,hes16,dfh+16}) that it has a spherically symmetric 
morphology centered at the GC, and the Navarro-Frenk-White (NFW) profile
was generally used in previous studies. However, other profiles such as
a generalized NFW (g-NFW) template and templates obtained from different 
considerations were also used, and the spectral results were found to
be dependent on the chosen profiles (e.g., \citealt{aaa+17}).
In the analysis we first adopted a simple geometric spatial distribution 
for the GCE emission to reduce the dependence of the results on the template 
profiles. Considering most of the GCE emission is detected within 
10$^{\circ}$ of the GC (see Figures 19 and 20 in \citealt{aaa+17}), 
we set 1 uniform disk and 9 rings between 0$^{\circ}$ 
and 10$^{\circ}$ from the GC with a radius step of 1$^{\circ}$; in other words,
10 GCE components were set, with radius 0$^{\circ}$--1$^{\circ}$,
1$^{\circ}$--2$^{\circ}$, ..., and 9$^{\circ}$--10$^{\circ}$,
denoted as GCE$_{1}$, GCE$_{2}$, ..., and GCE$_{10}$ respectively.

\subsection{Maximum Likelihood Analysis}
\label{sec:la}

We performed the maximum likelihood analysis to the LAT data in the RoI 
using the source model set above. In the analysis, simple power laws were 
used to describe the \gr\ emissions of the \fermi\ bubbles and the 10 GCE 
components. 
The prefactor parameters for them were set free, and the power-law indices 
were fixed to 2. The normalizations of the Galactic and extragalactic diffuse 
components were also set as free parameters. For all the catalog sources 
included in the source model, the spectral parameters were fixed to 
the catalog values given in \citet{4fgl-updated4}. Following
\citet{ccw15}, we introduced an 
energy-dependent weight map of the RoI in the likelihood analysis to 
minimize the impact of the point sources in our source model.
The weights were defined as 
\begin{equation}
\label{equ:wa}
w_{i,j} = \frac{1}{(\frac{\mu_{i,j}^{PSC}}{f_{PSC}\mu_{i,j}^{BG}})^{\alpha_{PSC}} +1}, 
\end{equation}
where $\mu_{i,j}^{PSC}$ and $\mu_{i,j}^{BG}$ are the expected numbers of 
photons from the point sources and the Galactic diffuse emission in 
the $i^{th}$ energy bin and $j^{th}$ pixel, respectively, calculated by 
creating model maps in different energy bins using \textit{gtmodel} in 
{\tt Fermitools}. The two scale parameters $\alpha_{PSC}$ and $f_{PSC}$ 
were fixed to the default values of 5 and 0.1 \citep{ccw15}, respectively. 
In this way, the LAT events in the $[i,j]$ bins around brighter sources had
lower weights.

The maximum likelihood analysis was performed in 25 evenly divided energy 
bins in logarithm from 0.3 GeV to 500 GeV.  For the data points obtained
for the 10 GCE components, only those with the fluxes greater 
than the statistic uncertainties were kept.
We noted that the systematic uncertainties dominated in the uncertainties 
of the GCE spectral results. The uncertainties induced by the Galactic 
diffuse emission templates were commonly considered in previous studies, 
which were generally evaluated by adopting different templates 
created with the \textit{GALPROP} code 
(e.g., \citealt{aaa+12,ccw15,zlh+15,aaa+17}).
In the analysis, we evaluated the systematic uncertainties by repeating 
the likelihood analysis in each energy bin with the normalizations 
of the Galactic diffuse component artificially fixed to 
values $\pm$6\% deviating from the best-fit values. The deviations represent
the local departures from the best-fit diffuse model, and they were found 
to be $\sim$6\% when analyzing source-free regions on the Galactic 
plane \citep{abdo+w51c2009,abdo+w28-2010}.

\begin{figure*}
   \centering
   \includegraphics[width=0.45\textwidth]{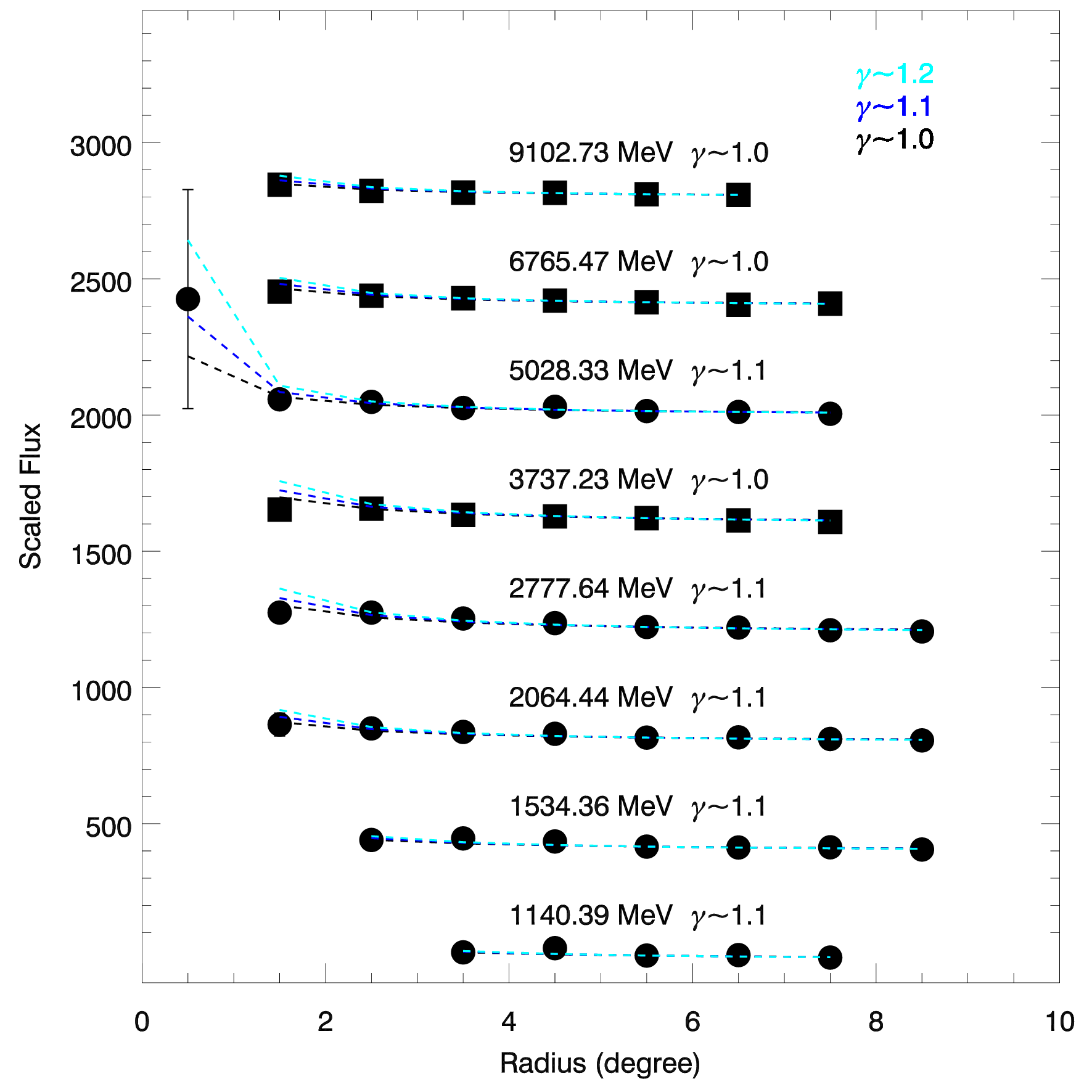}
    \includegraphics[width=0.45\textwidth]{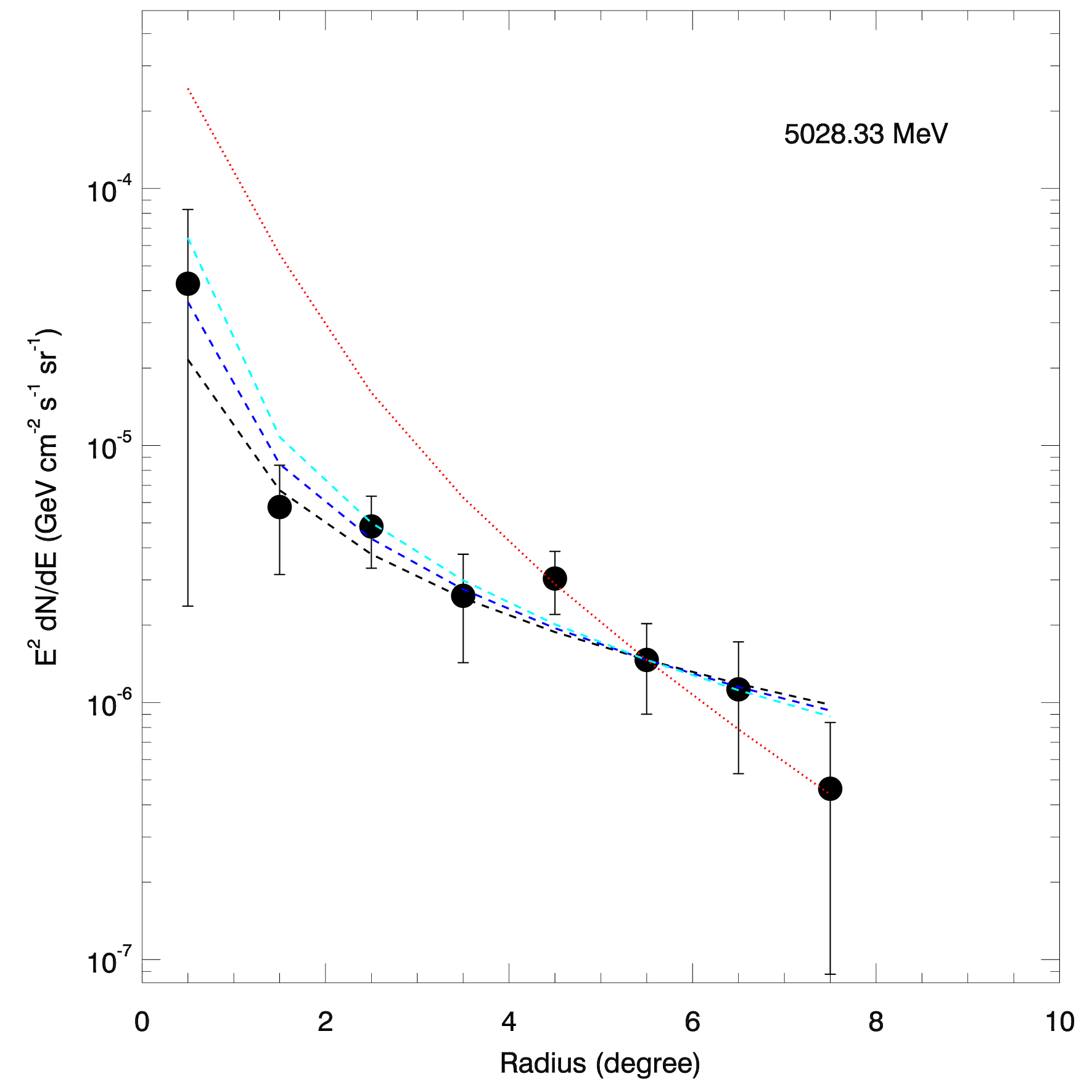}
   \caption{{\it Left}: rescaled \gr\ fluxes at different distances from 
	the GC. 
Eight energy bands centered from 1.1\,GeV to 9.1\,GeV were considered. 
The predicted fluxes from g-NFW profiles with $\gamma$ values of 1.0, 1.1, 
and 1.2 are plotted as dashed lines. The $\gamma$ value labeled with an 
energy band is the best-fit one for that energy band. 
{\it Right}: \gr\ fluxes at distances of 0\arcdeg --8\arcdeg\ from the GC 
at 5.0\,GeV, with the g-NFW profiles plotted as dashed lines (the same
as those in the left panel). The predicted profile from the stellar density 
	distribution is plotted as a red dotted line.  }
   \label{fig:profiles}
\end{figure*}

\begin{figure*}
   \centering
   \includegraphics[width=0.45\textwidth]{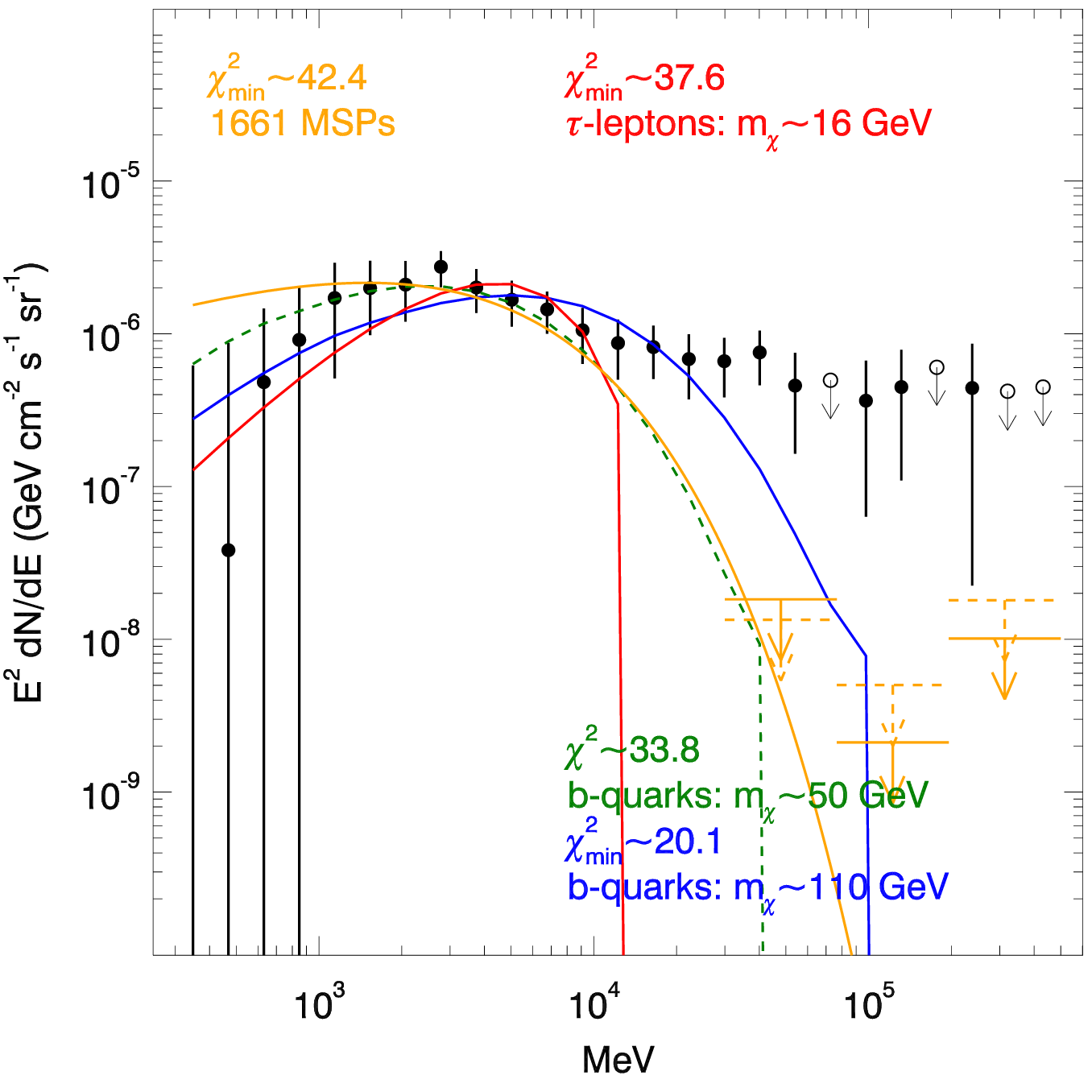}
   \includegraphics[width=0.45\textwidth]{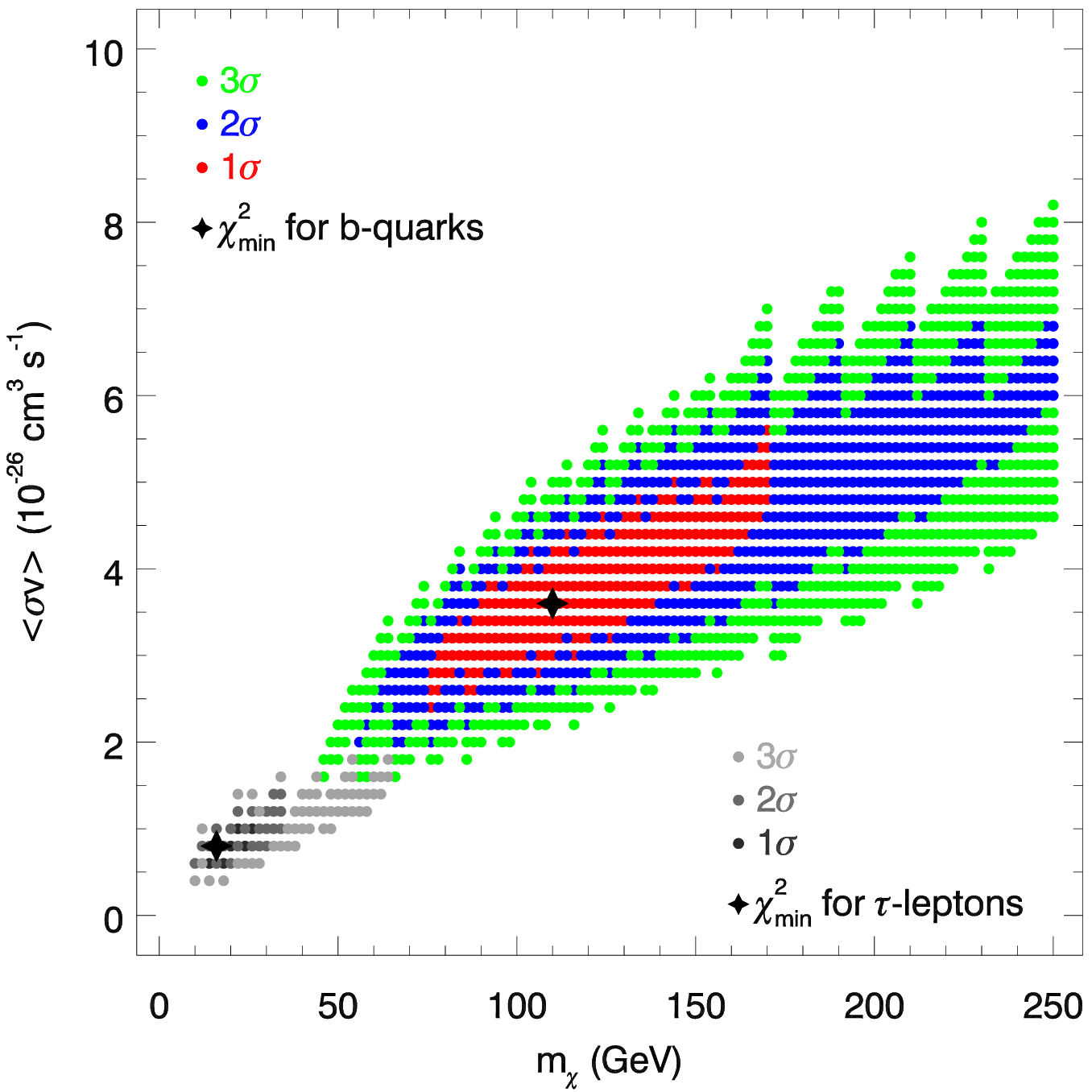}
   \caption{{\it Left}:  \gr\ spectrum of the GCE emission. The best-fit 
spectra from the MSPs, and those from the DM annihilating to $b\bar{b}$ and to 
$\tau^{+}\tau^{-}$ are plotted as yellow, blue and red lines, respectively. 
The model spectrum from the DM annihilating to $b\bar{b}$ with parameters 
adopted from \citet{ccw15} is plotted as a green dashed line for comparison.
The upper limits from the stacking analysis of the data of 96 and 48 \gr\
MSPs in 30--500\,GeV are shown as yellow solid and dashed arrows respectively
	(see Section~\ref{sec:stack}).
{\it Right}: 3$\sigma$ uncertainty ranges for 
	m$_{\chi}$ and ${\langle \sigma v \rangle}$ 
	obtained from fitting with the two annihilation channels, 
while the best-fit values are marked as the dark stars. }
   \label{fig:fitting}
\end{figure*}

GCE$_{1}$ (0$^{\circ}$--1$^{\circ}$ from the GC) is in the masked region,
and for GCE$_{10}$, no significant detections were found nearly across 
the full energy range. 
We considered 8 successive energy bands for the morphology study in the 
energy range of 1--10 GeV, during which the GCE emission has smaller 
systematic uncertainties induced by different factors \citep{ack+17}. 
The obtained fluxes, rescaled for clarity, are shown in 
the left panel of Figure~\ref{fig:profiles}. We 
obtained the best-fit 
g-NFW profile when $\gamma$ = 1.1 (see below Section~\ref{sec:ms} 
for details). We then adopted this profile for the full GCE in the 
likelihood analysis to extract the \gr\ spectrum using the procedures 
described above. The obtained spectrum for the full GCE emission is shown
in the left panel of Figure~\ref{fig:fitting}. For the spectral data points,
we kept those when the flux values are $\geq$2 times greater than their 
statistic uncertainties and otherwise used the derived 95\% flux upper limits. 
There were cases when the $-$6\% deviation from a best fit for the Galactic 
emission was set, the flux value turned to be $>$2 times greater 
than the uncertainty. For these cases, we replaced the upper limits
with the original flux measurements.

\section{Morphology of the GCE}
\label{sec:ms}

We studied the morphology of the GCE emission based on the g-NFW 
profile \citep{nfw96,nfw97}, 
\begin{equation}
\label{equ:nfw}
\rho(r)= \rho_{0}\frac{(r/r_{s})^{-\gamma}}{(1+r/r_{s})^{3-\gamma}}.
\end{equation}
A scale radius of $r_{s}$ = 20 kpc was adopted, and $\rho_{0}$ was 
selected such that the local DM density (at 8.5 kpc from the GC) is 
0.4\,GeV\,cm$^{-3}$ \citep{loc+11}.
We considered three profiles with $\gamma$ values of 1.0, 1.1, and 1.2 
respectively (see Figure~\ref{fig:profiles}), and fit the observed fluxes 
at different radii with the rescaled model fluxes (the model flux 
of a profile at 5.5$^{\circ}$ was set to be the observed value).
$\chi^2$ values were obtained from the fitting and the reduced $\chi^2$ values,
which are $\chi^2$ divided by the degree of freedom (DoF), were used 
to determine the best fits.
In our fitting, we noted that the model fluxes with $\gamma$=1.2 
at 1.5$^{\circ}$ were higher than the observed ones in six out of the eight 
energy bands centered at $\sim$2.1--9.1 GeV and particularly
in the energy band centered at $\sim$3.7\,GeV, a large $\chi^{2}$ value 
of 17.9 was obtained. Because of these, 
we did not further tested larger $\gamma$ values in the fitting.
We noted that the stellar density of the Galactic bulge given in \citet{vgg09} 
predicts a steeper slope than the g-NFW profile with $\gamma$=1.2.
An example is shown in the right panel of Figure~\ref{fig:profiles},
which are the observed fluxes (without being rescaled)
in the energy band centered at $\sim$5.0\,GeV. The predicted
profile from the stellar density distribution (red dotted line) significantly
deviates from the observed fluxes. Therefore this steeper profile was not
considered in our analysis.

The $\gamma$ values from the best-fits in each energy bands are provided in
Figure~\ref{fig:profiles}.
In three energy bands centered at $\sim$3.7, 6.8, and 9.1 GeV (plotted as 
squares in Figure~\ref{fig:profiles}), a $\gamma$ value of 1.0 was obtained 
with $\chi^{2}$ values of 6.1, 2.7, and 0.4 (for 6, 6, and 5 DoF), 
and in the other five energy bands (plotted as circles in 
Figure~\ref{fig:profiles}), a $\gamma$ value of 1.1 was obtained 
with $\chi^{2}$ values of of 2.0--6.3 (for 4--7 DoF).
There was only one significant detection for GCE$_{1}$, which was at 5.0 GeV, 
and $\gamma$=1.1 was obtained in this energy band. 

When we considered the g-NFW profile with a $\gamma$ value of 1.1
for the energy bands centered at 3.7, 6.8, and 9.1 GeV, the $\chi^{2}$ values
were 9.5, 3.3, and 1.7, also acceptable for describing the emission morphology
in these three bands comparing to those with a $\gamma$ value of 1.0. 
Therefore, we adopted  $\gamma$=1.1 for 
the g-NFW profile to describe the full GCE emission throughout the
rest of the paper.

\section{Spectral Fitting}
\label{sec:sf}

\subsection{Millisecond pulsars}
\label{sec:msp}

\begin{figure}
   \centering
   \includegraphics[width=0.45\textwidth]{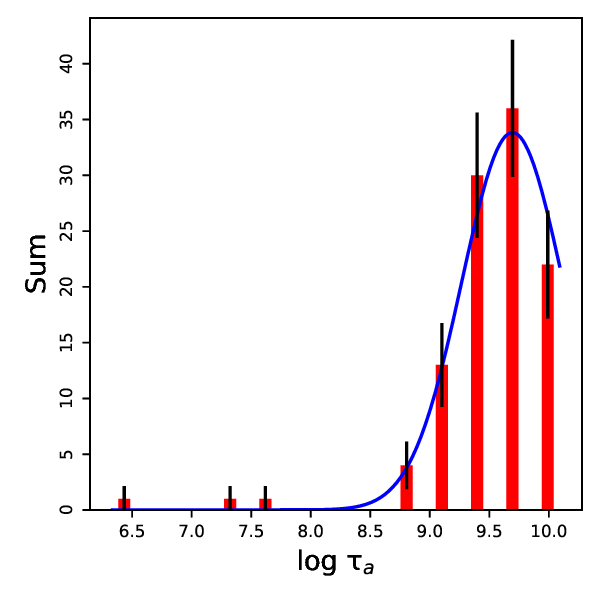}
   \caption{$\log(\tau_a)$ distribution of the known \gr\ MSPs and 
the corresponding best-fit log-normal function (the blue line).  }
   \label{fig:age}
\end{figure}

We followed the procedure given in \citet{wu+22} and tested to fit 
the full \gr\ spectrum of the GCE emission with the spectral templates 
constructed for varied numbers of putative MSPs in the GC region. 
The \gr\ spectra of the MSPs were generated based on the spectral
parameter (photon index $\Gamma$ and cutoff energy $E_c$) ranges
determined from the known \gr\ MSPs, where the distributions of their
spin periods ($P$) and characteristic ages ($\tau_a$), and the relation between 
their \gr\ efficiencies ($\eta$) and $\tau_a$ (see \citealt{wu+22} for details)
were considered. One difference in the fitting procedure was that 
we used $\tau_a$ of all known \gr\ MSPs to obtain the distribution in a form of
$\log(\tau_a)$, while in \citet{wu+22} because the targets were MSPs in
globular clusters, the $\tau_a$ distribution of MSPs was derived by comparing
with the ages of the host globular clusters.
In Figure~\ref{fig:age}, we show the $\tau_a$ distribution of the known \gr\ 
MSPs, which was fitted with a log-normal function 
\begin{equation}
p(\tau_a) = N e^{-\frac{1}{2\sigma^{2}}(\log\tau_a - \mu)^{2}}\ \ \ ,
\end{equation}
where $N$ is a normalization factor.
From fitting, we obtained $\mu = 9.69$ and $\sigma = 0.42$. This function
was used to generate $\tau_a$ for assumed MSPs.

We ran 1000 times of the fitting to obtain the best fit with
the minimum $\chi^{2}$ ($\chi^{2}_{min}$) and a range for numbers of MSPs, 
where a source distance of 8.5 kpc was adopted and the latter 
was estimated from the lowest 5\% $\chi^{2}_{min}$ 
values in the fitting runs. The best-fit number of MSPs was 
1661 with $\chi^{2}_{min}=42.4$
and the number range was 1415--1706.
These results are given in Table~\ref{tab:fitting}.

The best-fit spectrum, emitted from 1661 MSPs, is shown in 
Figure~\ref{fig:fitting}. As can be seen, it can approximately describe 
the low, $<$10\,GeV part of the spectrum of the GCE emission, but 
the high energy part can not be explained at all
since \gr\ MSPs generally have a spectral cutoff at $\sim$1--2\,GeV. 
In addition,
the low-end of the model spectrum appears to be higher than the observed 
data points.

\subsubsection{Stacking analysis of $\geq$30\,GeV emission from millisecond pulsars}
\label{sec:stack}

The MSP scenario for the GCE suffers different problems (e.g.,
\citealt{hm16}), and one obvious is the lack of high-energy emission
from MSPs, as shown in Figure~\ref{fig:fitting}. We investigated this problem 
with the stacking technique.

\begin{figure}
   \centering
   \includegraphics[width=0.40\textwidth]{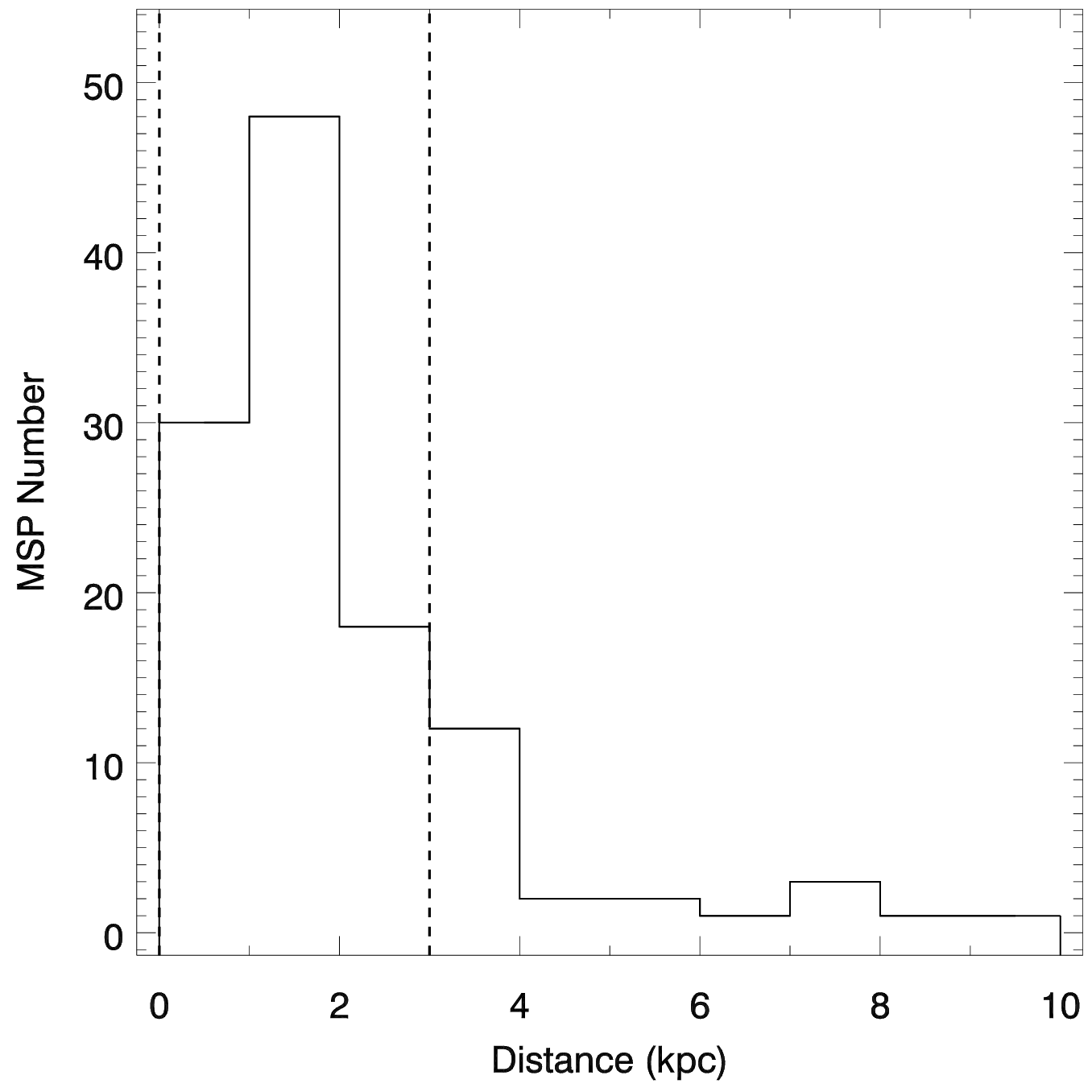}
   \caption{Distance distribution of 115 known \gr\ MSPs. 
	The range of 0--3 kpc is marked by the dashed lines.
}
   \label{fig:d}
\end{figure}

Whether pulsars have high-energy emission upto TeV energies has been probed and
the very recent detection of pulsed TeV emission from the Vela 
pulsar has seemingly provided encouraging results \citep{hess-vela}.
\gr\ spectra of 104 MSPs in \citet{wu+22} were obtained, and only one MSP, 
PSR~J0614$-$3329, was noted to have significant detectable \gr\ emission
in $>$30 GeV energy range (see Figure 2 in \citealt{wu+22}). Here we 
applied a stacking technique to the analysis of the 30--500 GeV LAT data 
of \gr\ MSPs, for the purpose of finding if MSPs could have any magnetospheric 
emission that possibly
match the high-energy part of the GCE. There are 134 \gr\ MSPs in the Galaxy 
listed in 4FGL-DR4, among which 115 have distance values reported in 
the ATNF pulsar catalog \citep{man+05}. The distance distribution of them 
is shown in Figure~\ref{fig:d}. There are 48 MSPs within
distances of 1--2\,kpc and 96 MSPs within distances 
of 0--3\,kpc. We selected the 96 MSPs with distances 
$\leq$3\,kpc as the target sources. 
The same time period of the LAT data as that for the GCE was chosen, 
but the \textit{SOURCE} event class of the data was used, which is 
recommended for studies of point sources.

We first performed the binned likelihood analysis to the 30--500 GeV LAT 
data within a $20\arcdeg\times 20\arcdeg$ region centered at each of
the 96 MSPs. For each target MSP, a source model containing all the catalog 
sources within 20\,deg from the MSP was used. The background Galactic 
and extragalactic diffuse 
spectral models (gll\_iem\_v07.fits and iso\_P8R3\_SOURCE\_V3\_v1.txt
respectively) were also included in this source model. 
The normalizations of the catalog sources within 5\,deg from the target 
and the two background diffuse emission models were set as free parameters.
From the analysis, we again found that only PSR\,J0614$-$3329 could be 
significantly detected in the 30--500\,GeV band, having a Test Statistic (TS) 
value of 46. The 
second high TS value was only $\sim$6, obtained for PSR J1311$-$3430.

We then performed the stacking analysis of the 30--500 GeV data of the 96 MSPs.
The procedure of the stacking analysis, partly referring to that described 
in \citet{hub+12}, is as follows:
\begin{enumerate}
\item For each of the targets, we extracted a model count cube 
	in 10 evenly divided energy subbands in logarithm in the interested 
	energy band, by running \textit{gtmodel} in {\tt Fermitools} in 
	each of the subbands.  The catalog sources included in the 4FGL-DR4 
	were considered in the source model, with the spectral parameters 
	fixed at the values obtained from the above likelihood analysis. 
\item We extracted the residual count cube for each of the targets 
	by subtracting the model one from the observed one.
\item The residual count cubes for all targets were added together for 
	the following analysis (at step 6).
\item For each of the targets, we extracted the source count cube in the 10 
	energy subbands with a source model including only the target source 
	and the background Galactic and extragalactic diffuse emission. 
	The source count cube was obtained by running \textit{gtsrcmaps} 
	in {\tt Fermitools}, and the exposure cube (obtained by running 
	\textit{gtexposure}) was considered.
\item The source count cubes for all targets were added together as a final 
	source count cube for the following analysis.
\item The likelihood analysis was performed to the added residual count cube 
	with the final source count cube. The \gr\ emission from the added 
	target was described with a power law with $\Gamma$ fixed at 2. 
	The normalizations of the target and the two background emissions 
	were set free.
\end{enumerate}

\begin{figure*}
   \centering
   \includegraphics[width=0.45\textwidth]{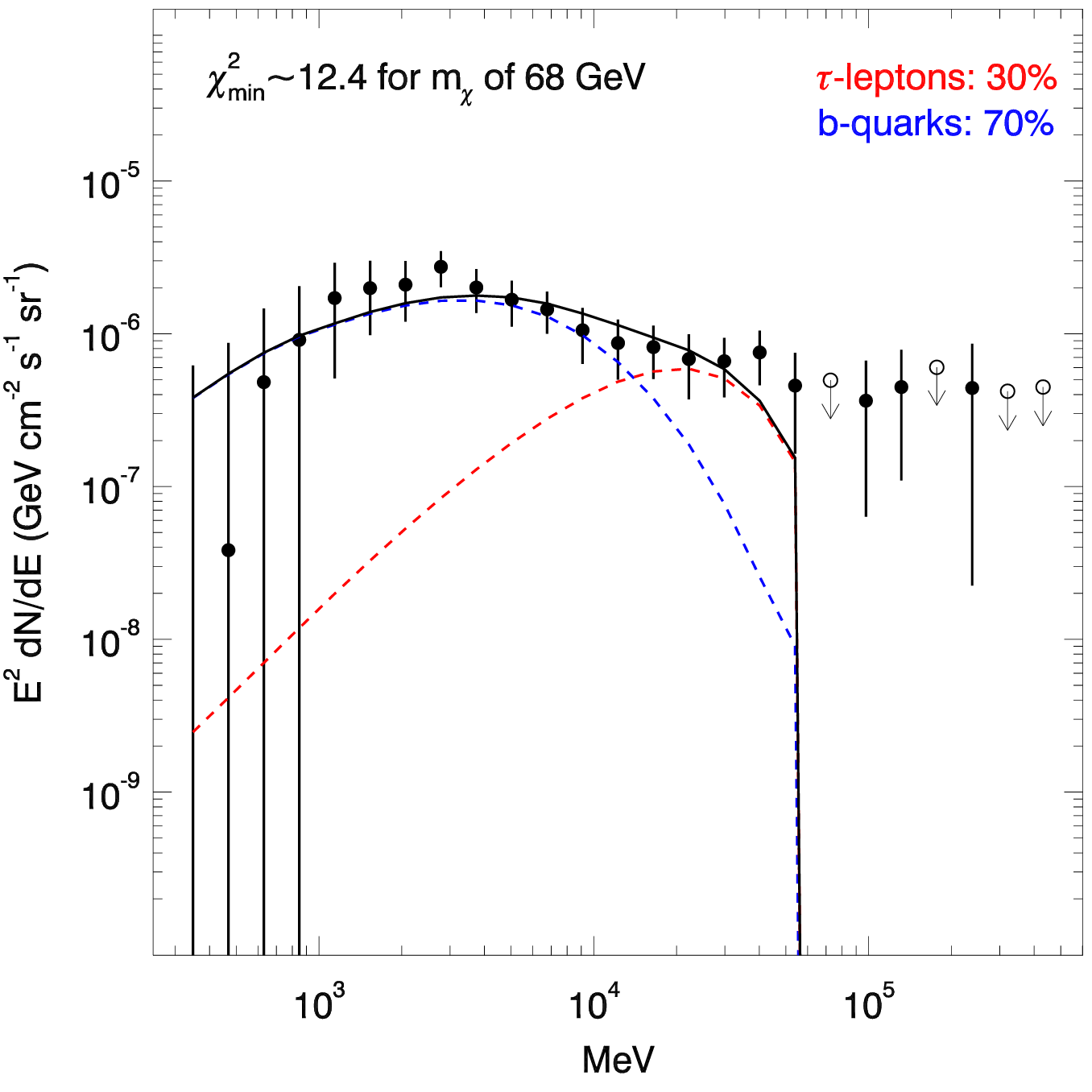}
   \includegraphics[width=0.45\textwidth]{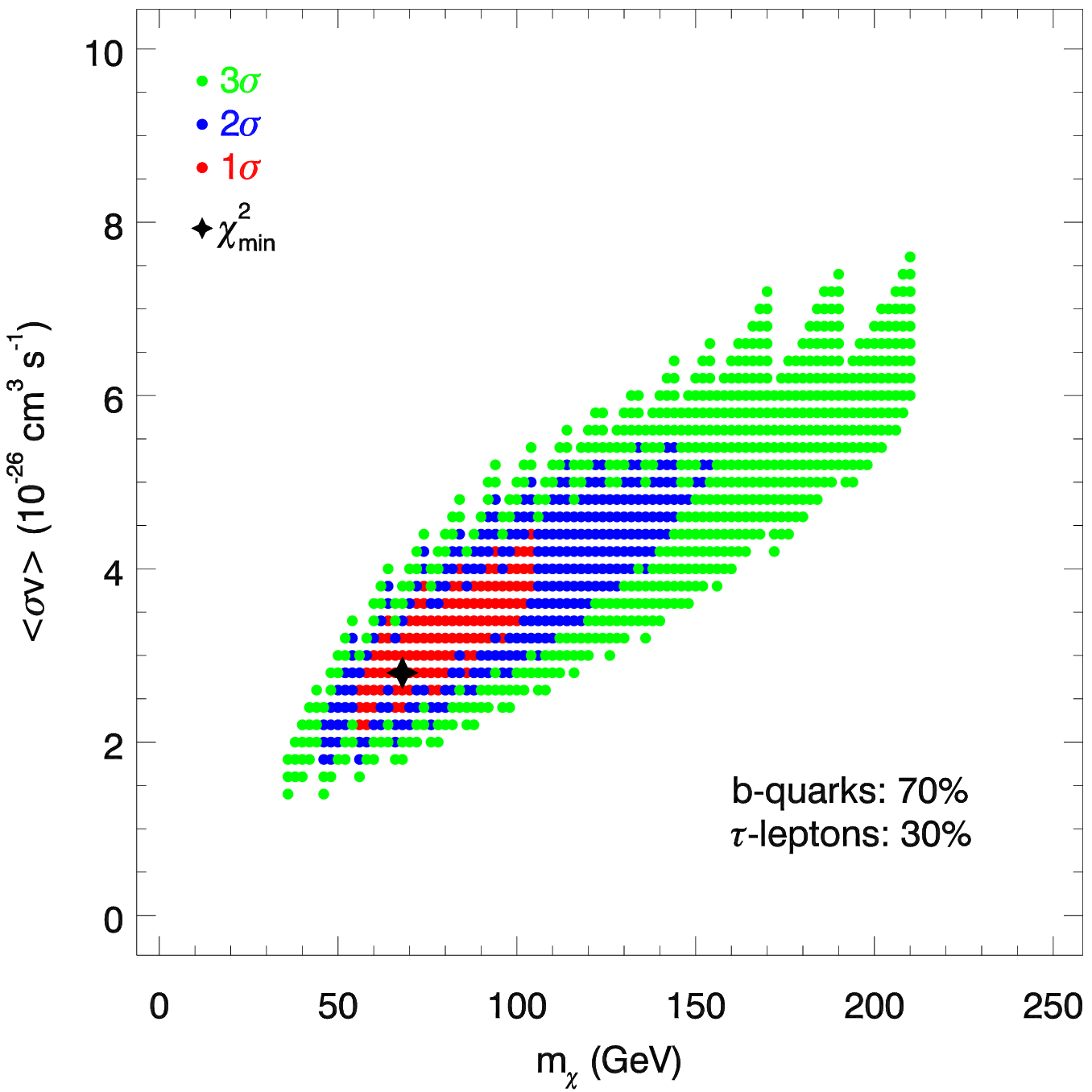}
   \caption{Fitting to the \gr\ spectrum of the GCE emission
with a combination of the DM annihilating to $b\bar{b}$ and to 
	$\tau^{+}\tau^{-}$.
	  {\it Left}: the branching ratio into $b\bar{b}$ of 0.7
	  provides the best fit (black solid line), to which the $b\bar{b}$
	  and $\tau^{+}\tau^{-}$ contributions are indicated by the blue 
	  and red lines respectively. 
{\it Right}: 3$\sigma$ uncertainty ranges for m$_{\chi}$ and 
	${\langle \sigma v \rangle}$, 
	with the best-fit values marked as a black star.  }
   \label{fig:fitting_2dm}
\end{figure*}

In the analysis, we repeated the stacking analysis in 3 energy bands divided 
in logarithm in 30--500 GeV, and derived the 95\% upper limits for the added 
target in each of the bands. The results are given in 
Table~\ref{tab:stacking}. In the first energy band of 30.0--76.6 GeV,
a TS value of 14 was obtained, corresponding to a detection significance 
of $>$3$\sigma$. This detection was likely due to emission of PSR J0614$-$3329
in this energy band.  To compare with the GCE emission, we calculated 
the predicted flux upper limits for 1661 MSPs 
located at 8.5\,kpc, assuming them within the spatial extent of the GCE. 
For the calculation, the added target was assumed at a distance of 1.5\,kpc, 
since we only 
analyzed the data for the MSPs with distances of 0--3 kpc and half of them
have distances of 1--2 kpc. The obtained upper limits are plotted in 
Figure~\ref{fig:fitting} (the yellow solid arrows). As can be seen,
the spectral data points of the GCE in the high energy range are higher 
than the upper limits.

We tested to include only the 48 \gr\ MSPs with distances of 1--2 kpc 
for the stacking analysis to check any possible effect due to the relatively
large MSP distance range. The TS value for the added source in the first 
energy band 
(30.0--76.6 GeV) was only $\sim$2 (since PSR J0614$-$3329 has a distance of
0.6 kpc and was not included). The obtained upper limits are shown as
the dashed yellow arrows in Figure~\ref{fig:fitting}, still
lower than the spectral data points of the GCE.
We concluded that MSPs alone likely cannot explain the GCE emission.
\begin{figure}
   \centering
   \includegraphics[width=0.4\textwidth]{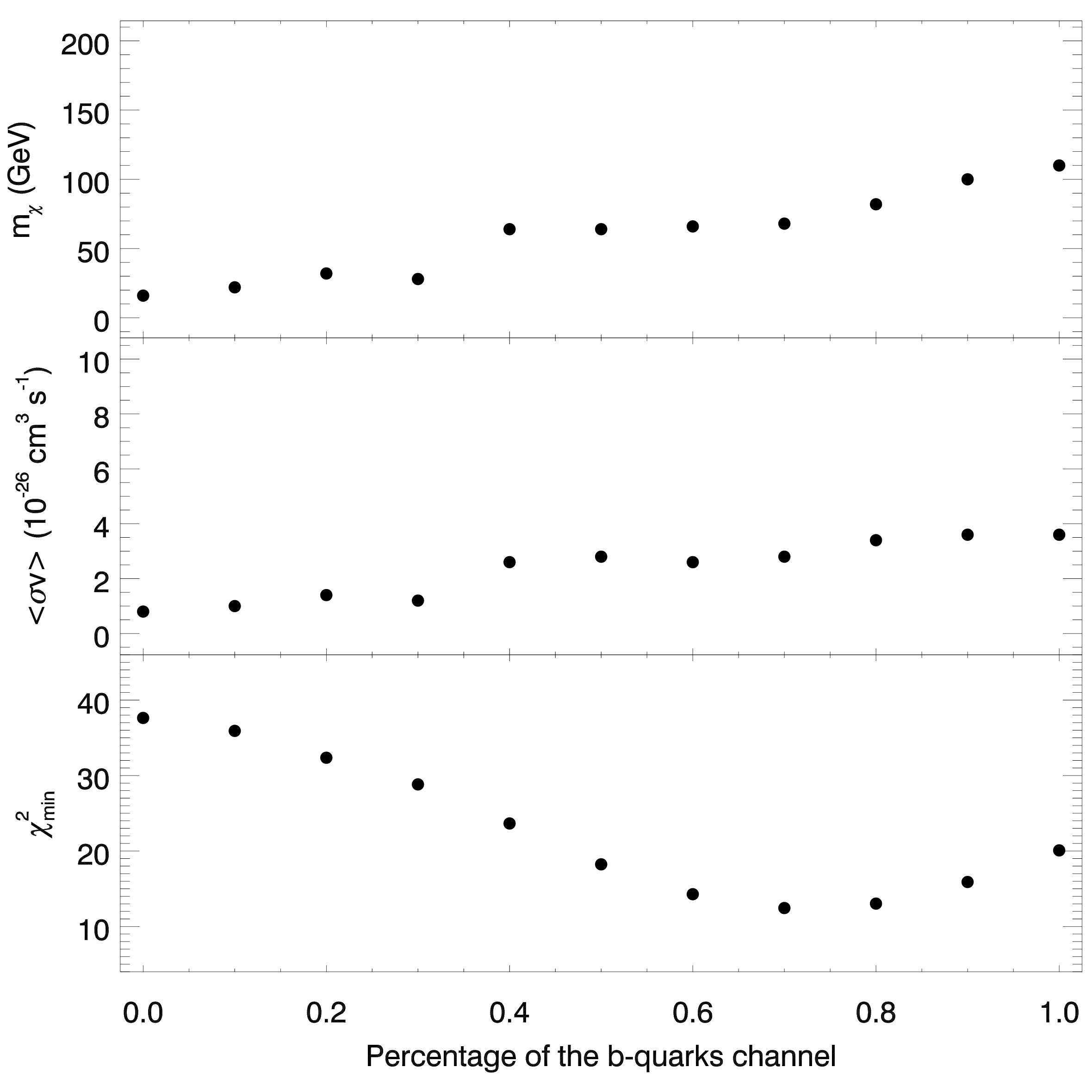}
   \caption{Obtained $\chi^{2}_{min}$ values as a function the branching 
	ratio into $b\bar{b}$ (bottom panel). The corresponding
	${\langle \sigma v \rangle}$ and m$_{\chi}$ values 
	are shown in the middle and top panel respectively.  }
   \label{fig:chi2_2dm}
\end{figure}

\subsection{Dark matter}
\label{sec:dm}

In order for a comparison, we considered the DM annihilation process to fit 
the \gr\ spectrum of the GCE emission. The flux of the \gr\ emission 
generated by DM annihilations 
is given by
\begin{equation}
\label{equ:spectra}
\phi(E_{\gamma},\phi)=\frac{{\langle \sigma v \rangle}}{8\pi m^{2}_{\chi}}\frac{dN_{\gamma}}{dE_{\gamma}}\int_{los}\rho^{2}(r)dl,
\end{equation}
where $m_{\chi}$ is the DM mass, ${\langle \sigma v \rangle}$ is 
the annihilation cross section, $dN_{\gamma}/dE_{\gamma}$ is the \gr\ spectrum
produced per DM annihilation, and
$\rho(r)$ is the DM density as a function of distance $r$ to the GC with
the integral calculated over the line-of-sight (los).
In this fitting, a $m_{\chi}$ range of 2--250\,GeV with a step of 2\,GeV and 
a ${\langle \sigma v \rangle}$ range of 0.2--10$\times$10$^{-26}$ cm$^{3}$\,s$^{-1}$ 
with a step of 0.2$\times$10$^{-26}$ cm$^{3}$\,s$^{-1}$ were searched.
The $dN_{\gamma}/dE_{\gamma}$ spectra were generated using the 
\textit{gammapy.astro.darkmatter} module in \textit{Gammapy v0.20},
which is based on \citet{cir+11} and provides tabulated spectral values
for different annihilation channels. As the annihilation channels 
$\tau^{+}\tau^{-}$ and $b\bar{b}$ were commonly considered in previous 
studies (e.g., \citealt{hg11,ak12,dfh+16,hes16,d21}), we also adopted these
two channels.

We found that for the channel $b\bar{b}$, a $m_{\chi}$=110 GeV DM with  
${\langle \sigma v \rangle}$ = 3.6$\times$10$^{-26}$\,cm$^{3}$\,s$^{-1}$ 
provides the model spectrum that relatively well fits that of 
the GCE emission, for which $\chi^{2}_{min}$=20.1.
For the channel $\tau^{+}\tau^{-}$, a $m_{\chi}$=16 GeV DM with 
${\langle \sigma v \rangle}$ = 0.8$\times$10$^{-26}$\,cm$^{3}$\,s$^{-1}$ 
provides the best-fit spectrum among those from the channel,
but with a high $\chi^{2}_{min}$ value of 37.6.
These fitting results are given in Table~\ref{tab:fitting}, and the
best-fit model spectra are shown in the left panel of Figure~\ref{fig:fitting}. 
In addition, the 1$\sigma$--3$\sigma$ uncertainty ranges for 
the two parameters are provided in the right panel of 
Figure~\ref{fig:fitting}. 

A combination of the channels $b\bar{b}$ and $\tau^{+}\tau^{-}$ was tested
by us, where a range of 0.0--1.0 for the branching ratio into $b\bar{b}$ 
was set. The best fitting was found when the branching ratio was 0.7,
for which the model spectra are shown in the left panel of
Figure~\ref{fig:fitting_2dm}.  In Figure~\ref{fig:chi2_2dm}, the
resulting $\chi^{2}_{min}$ values, as well as those of the DM's $m_{\chi}$ 
and ${\langle \sigma v \rangle}$, are shown. The minimum $\chi^{2}_{min}$ 
was 12.4,
when $m_{\chi}$ and ${\langle \sigma v \rangle}$ were 68\,GeV and 
2.8$\times$10$^{-26}$ cm$^{3}$\,s$^{-1}$
respectively (also given in Table~\ref{tab:fitting}). 
The 1$\sigma$--3$\sigma$ uncertainty ranges for the 
parameters are shown in the right panel of Figure~\ref{fig:fitting_2dm}. 

\subsection{Joint fitting}
\label{sec:msp-dm}

\begin{figure}
   \centering
   \includegraphics[width=0.45\textwidth]{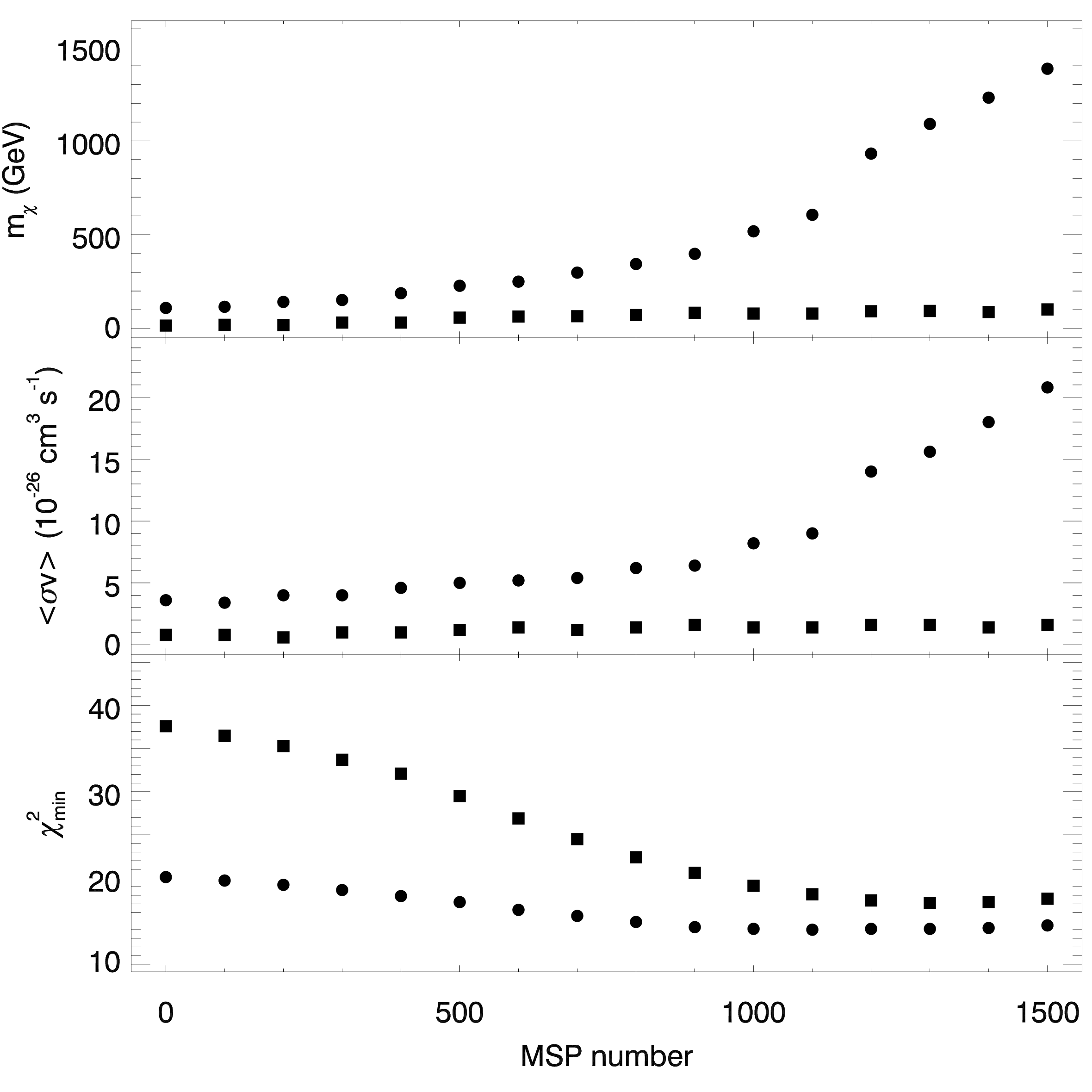}
   \caption{$\chi^{2}_{min}$ values resulting from different MSP numbers
	(bottom panel),
	obtained with fitting the GCE \gr\ spectrum with those of the MSPs 
	plus DM annihilating to $b\bar{b}$ (circles) or to $\tau^{+}\tau^{-}$ 
	(squares). The corresponding DM's ${\langle \sigma v \rangle}$ and 
	m$_{\chi}$ values are shown in the middle and top panel respectively. }
   \label{fig:fitting_msp_dm}
\end{figure}

We also explored the possibility of the contributions of both MSPs and DM 
annihilations to the GCE.  Since the lower limit for 
the MSP numbers estimated above was 1415, we set the MSP number from 100 
to 1500 with a step of 100. Same as the above in Section~\ref{sec:dm},
each of the annihilation channels $\tau^{+}\tau^{-}$ and $b\bar{b}$ was 
considered. 
In the fitting, we found that the channel $b\bar{b}$ often required 
large $m_{\chi}$ and ${\langle \sigma v \rangle}$ values when the number of 
MSPs was increased, and thus we considered ranges of 2--2000\,GeV and 
0.2--50$\times$10$^{-26}$\,cm$^{3}$\,s$^{-1}$ (wider than above in 
Section~\ref{sec:dm}) for $m_{\chi}$ and ${\langle \sigma v \rangle}$ 
respectively.  For each-set number of MSPs, 1000 runs were conducted. The
resulting 1000 spectra were averaged to obtain a single model spectrum, with 
the standard deviations of the former minus the latter being
the uncertainties.
The resulting $\chi^{2}_{min}$ values, as well as the DM's $m_{\chi}$ 
and ${\langle \sigma v \rangle}$ values, are shown in 
Figure~\ref{fig:fitting_msp_dm}, with the best-fit parameter results 
($\chi^{2}_{min}$ = 14.0 and 17.1 respectively for the channels $b\bar{b}$ and 
$\tau^{+}\tau^{-}$) given in Table~\ref{tab:fitting}. 

\begin{figure*}
   \centering
   \includegraphics[width=0.45\textwidth]{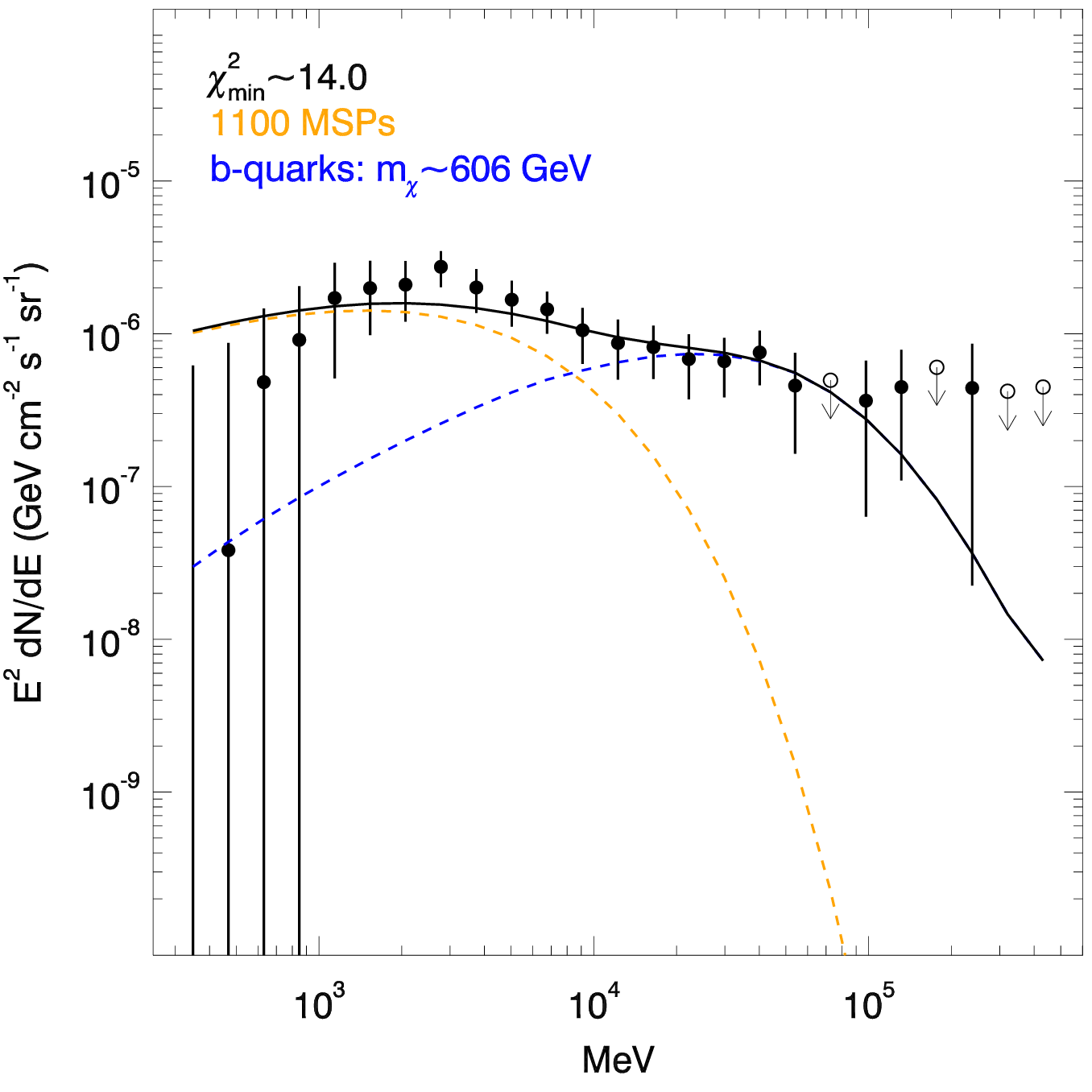}
   \includegraphics[width=0.45\textwidth]{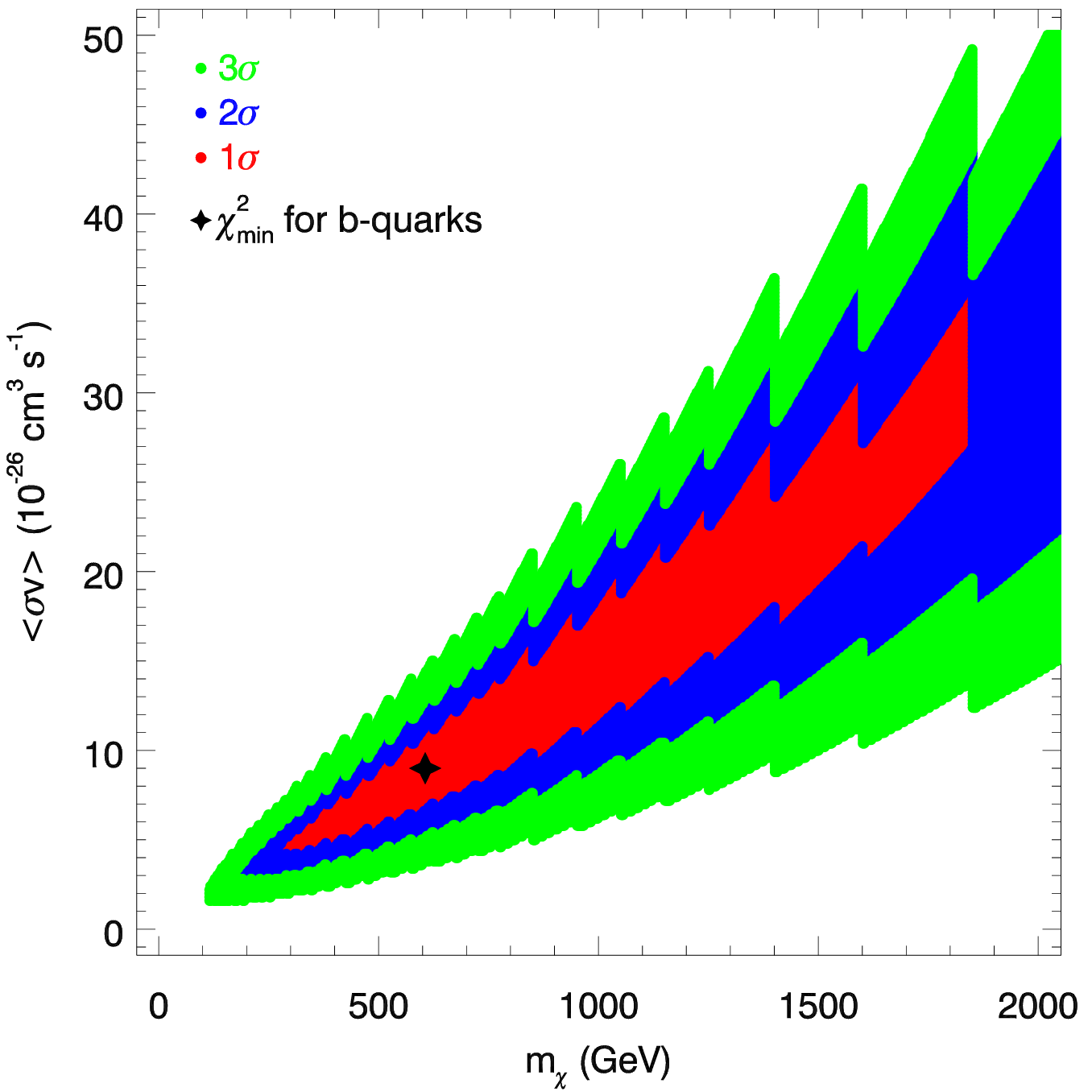}
   \caption{MSP plus the DM channel $b\bar{b}$ fitting to the \gr\ spectrum
of the GCE emission. {\it Left}: the best fit provided by 1100 MSPs 
	(yellow dashed line) plus the DM's m$_{\chi}$=606\,GeV and 
	${\langle \sigma v \rangle}$ = 9.0$\times 10^{-26}$\,cm$^3$\,s$^{-1}$
	(blue dashed line).  The spectrum added from the two components is 
	shown as a black solid line.  {\it Right}: the 3$\sigma$ uncertainty 
	ranges for the parameters of the DM channel $b\bar{b}$, with the
	best-fit values marked as a black star. }
   \label{fig:spectra_msp_dm}
\end{figure*}

We found that the MSP plus $b\bar{b}$ model always provided better fits 
than the MSP plus $\tau^{+}\tau^{-}$ model (see Figure~\ref{fig:fitting_msp_dm})
for a given number of MSPs. Given that the pure DM annihilation model through 
the channel $b\bar{b}$ also provided better fits than that through 
the channel $\tau^{+}\tau^{-}$ (Section~\ref{sec:dm}), we focused on 
the details of the MSP plus $b\bar{b}$ model fitting. 
The spectral fitting results from the 1100 MSPs plus $b\bar{b}$ model is 
shown in the left panel of Figure~\ref{fig:spectra_msp_dm}, which had
the minimum $\chi^{2}_{min}$ value (14.0) in the joint fitting. The 
1$\sigma$--3$\sigma$ uncertainty ranges for the DM's parameters are shown 
in the right panel of Figure~\ref{fig:spectra_msp_dm}.

\section{Discussion and Summary}
\label{sec:dis}

We have performed a detailed study of the GCE \gr\ emission by 
analyzing 15 yrs of the data collected with \fermi-LAT. 
The GCE region we considered was a circular one with radius 10\arcdeg. Among 25 energy bands divided from 0.3\,GeV 
to 500\,GeV, we obtained the flux measurements in 8 successive bands centered 
at 1--10\,GeV, which likely were the more significant part of the GCE emission.
The spatially resolved flux measurements were fit with a g-NFW profile, which
is often considered as that for the DM distribution in the GC, and $\gamma$=1.1
was obtained. This value is the same as that obtained in \citet{aba+14}, but
lower than the values of $\sim$1.2--1.3 obtained in other previous studies 
(e.g., \citealt{hg11,hl11,ak12,gm13,ccw15,cal+15,day+16,d21}). In our results,
$\gamma$=1.2 was marginally disfavored in the energy band centered at 3.7\,GeV
at a confidence level of $\sim$2.7$\sigma$ (a $\chi^{2}$ value of 17.9 was 
obtained for 6 DoF; see Section~\ref{sec:ms}).
As \citet{d21} have obtained a 10\% systematic uncertainty on $\gamma$, which
is 1.08--1.32, our result of $\gamma$=1.1 is within the uncertainty. 

Setting the g-NFW profile with $\gamma$=1.1 as a template in the source model, 
the \gr\ spectrum of the GCE 
emission was extracted. We would like to point out that this spectrum can
be affected by the profile chosen to be used, but the spectral shape is less
affected.  We tested the g-NFW profiles with $\gamma$ values of 1.0 or 1.2,
as well as other spatial templates such as the Moore \citep{dms04} and the 
Einasto \citep{mer+06,nav+10} profiles, and 
the resulting \gr\ spectra appeared higher or lower than that we obtained
with $\gamma$=1.1. However our comparison indicated that the spectral shapes 
were consistent with each other within uncertainties.

With the obtained spectrum of the GCE emission, we tested the MSP 
scenario by fitting it with those constructed from different numbers of MSPs. 
Previously,
a number of $\sim$10$^{3}$ MSPs was suggested to constitute a reasonable source
for the GCE (e.g., \citealt{ak12,gm13}). Our fitting limited the number to 
$\sim$1700. However the MSP-model spectrum suffers the problems of having
higher fluxes in the $\leq$1\,GeV energy range and lower fluxes in 
the $\geq$10\,GeV energy range comparing to the GCE's spectrum 
(Figure~\ref{fig:fitting}).
The minimum $\chi^{2}_{min}$ value we found was 42.4, even higher than that from
the fitting with the DM channel $\tau^{+}\tau^{-}$.
There are 21 spectral data points in the GCE spectrum (not including
the upper limits), and the generation of a spectrum for a number of MSPs
is the Monte Carlo process have six parameters ($P$, $\tau_a$, $\eta$, and
$\Gamma$ and $E_c$ of a \gr\ spectrum, plus
the number of MSPs). DoF in the fitting was 15, and 
$\chi^{2}_{min}$ = 42.4 corresponds to a $p$-value of 
$\sim$2$\times 10^{-4}$. This $p$-value suggests that the MSPs origin 
is rejected at a 3.7$\sigma$ confidence level.

The high-energy tail in the spectrum of the GCE emission has been noted in 
previous studies such as \citet{cal+15} and \citet{lin+16}. \citet{ccw15} 
pointed out that the use of the P6\_V11 Galactic diffuse emission model 
may lead to an over-subtraction of the GCE emission above $\sim$10 GeV. 
Our analysis with the model used still indicated significant detection 
of the high-energy tail. One possible explanation for the tail involves the 
scenario that electrons produced by pulsars ICS 
cosmic background microwave photons to high-energy ones \citep{lin+16}. 
This possibility has been used to interpret the high-energy \gr\ emission 
from the Sagittarius dwarf spheroidal galaxy \citep{cro+22}. However 
the ICS would be weak in an environment like the Galactic bulge because 
much of the energy carried by escaping electrons would be lost due to 
synchrotron 
radiation \citep{cro+22}. Because of the obvious discrepancy between
the GCE's and MSPs' spectra in $\geq$10\,GeV energy range, 
we also performed the stacking analysis of the 30--500 GeV data of 96 \gr\ 
MSPs for possibly resolving the discrepancy by checking if MSPs would have
significant high-energy emission. Only upper limits were 
obtained, and the
scaled values for 1661 MSPs were lower than the spectral data 
points of the GCE. Thus both the fitting results and the obvious discrepancy 
in the high energy range suggest that MSPs alone are not likely the source 
of the GCE emission.

We considered the DM annihilation models for the GCE emission for comparison, 
and found that the channel $b\bar{b}$ with $m_{\chi}$=110\,GeV and 
${\langle \sigma v \rangle}$ = 3.6$\times$10$^{-26}$\,cm$^{3}$\,s$^{-1}$ could
provide the best fit. The $\chi^{2}_{min}$ value was 20.1, corresponding to 
a $p$-value of 0.4 (for 19 DoF). The $m_{\chi}$ value is higher than that 
obtained in previous studies 
(several tens of GeV, e.g., \citealt{hg11,ak12,ccw15,cal+15,dfh+16,hes16,d21}).
We note that \citet{cal+15} found a maximum $m_{\chi}$ of $\sim$74 GeV 
at a $p$-value of $>$0.05 in the case of the channel $b\bar{b}$, which is 
within the uncertainty ranges we obtained (see the right panel of 
Figure~\ref{fig:fitting}). 
In Figure~\ref{fig:fitting}, a model spectrum with  $m_{\chi}$=50\,GeV
(${\langle \sigma v \rangle}$ = 1.8$\times$10$^{-26}$\,cm$^{3}$\,s$^{-1}$; see
\citealt{cal+15,ccw15}) is shown as a comparison example. As can be seen, 
this spectrum is very similar to our MSP model spectrum 
and can not provide any fits to the high-energy part of the GCE's spectrum
(the fitting gives $\chi^{2}_{min}$=33.8). 
There are differences between our analysis and previous ones that should 
be noted:
more LAT data and the latest source catalog were used in our analysis.

In order to fit the whole spectrum of the GCE emission as much as possible, 
the combination 
of the channels $b\bar{b}$ and $\tau^{+}\tau^{-}$ was tested.
The best fit had $\chi^{2}_{min}$ = 12.4,
corresponding to a $p$-value of 0.8 (for 18 DoF).
The branching ratio into $b\bar{b}$ was 0.7 for this case,
and the corresponding $m_{\chi}$ and ${\langle \sigma v \rangle}$ were
68\,GeV and 2.8$\times$10$^{-26}$\,cm$^{3}$\,s$^{-1}$ respectively.
The $m_{\chi}$ value is close to that obtained in the previously studies.
We applied the $F$-test to evaluation of the significance of an additional 
annihilation channel \citep{pro+02}. For an additional channel $b\bar{b}$ 
to the only channel $\tau^{+}\tau^{-}$, the $F$-test probability of finding 
a fit improvement by chance is 1$\times$10$^{-5}$, which corresponds to 
a significance of 4.4$\sigma$. For an additional channel $\tau^{+}\tau^{-}$ 
to the only channel $b\bar{b}$, the $F$-test probability is 
4$\times$10$^{-3}$, which corresponds to a significance of 2.8$\sigma$.

We also explored the case that both MSPs and DM annihilations contribute to 
the GCE emission. Given the low-energy cutoff in the \gr\ spectra of MSPs, 
the DM annihilation is thus needed for the high-energy part of the GCE emission.
Varying MSP number, the best-fit we obtained was when there were 1100 MSPs plus
the channel $b\bar{b}$ with $m_{\chi}\simeq 600$\,GeV and 
${\langle \sigma v \rangle}$ = 9.0$\times 10^{-26}$\,cm$^{3}$\,s$^{-1}$
($\chi^{2}_{min}$ = 14.0).
We note that in the joint model fitting, the
DM's $m_{\chi}$ and ${\langle \sigma v \rangle}$ were obtained with large 
uncertainties (see the right panel of Figure~\ref{fig:spectra_msp_dm}), which
is likely due to the large uncertainties of the spectral data points in 
the high energy range (i.e., can not provide strong constraints in the fitting).
While the MSPs essentially provide the low-energy part for the GCE emission
in the joint fitting,
there has also been the pure DM annihilation scenario suggested to
be able to provide a fit. In this scenario, a two-component DM model is 
required (e.g., \citealt{lhy11,aba+15}): the low-energy 
emission would arise from ICS of the electrons produced from the annihilations
of light DM particles. 

We estimate the fit improvement from the joint model to check if the MSPs are
needed. As in the fitting, the model spectrum of a given number of MSPs was
obtained by averaging 1000 spectra, with each resulting from a run,
the number of MSPs was the only additional parameter in addition to the two
parameters
from the DM annihilations.  Applying the $F$-test to the case of the MSPs 
plus the DM channel $b\bar{b}$ comparing to the pure DM channel $b\bar{b}$, 
the probability of finding a fit improvement by chance is $\sim$0.01, 
corresponding to a confidence level 
of $\sim$2.5$\sigma$. Thus an additional MSP component is only marginally 
needed for the GCE emission.

We summarize our results as follows:
\begin{enumerate}
\item A g-NFW profile with $\gamma$ of 1.1 is found to describe the \gr\ 
	morphology of the GCE.
\item Fitting the obtained spectrum of the GCE emission, $\sim$1700 MSPs are
	found to be needed, while the MSP model spectrum can not provide any 
		fits for the $\geq$10\,GeV part of the spectrum.
\item Stacking analysis is performed for 96 nearby \gr\ MSPs, but no 
	significant emission
	in 30--500\,GeV energy range is detected, which verifies the mis-match
		between the spectra of the GCE emission and the MSPs in the
		$\geq 10$\,GeV high-energy range.
\item DM annihilation channels $\tau^{+}\tau^{-}$ and $b\bar{b}$ are tested
	for fitting the GCE's spectrum. The latter provide a better fit than
	the former (with $p$-values of 0.4 and 6$\times$10$^{-3}$ respectively).
	The best-fit $m_{\chi}$ and ${\langle \sigma v \rangle}$ values of the 
	DM $b\bar{b}$ model are 110\,GeV and 
	3.6$\times$10$^{-26}$ cm$^{3}$\,s$^{-1}$, respectively. When
	combining the two channels, 70\% $b\bar{b}$ plus 
		30\% $\tau^{+}\tau^{-}$ provides the best fit. 
\item For the case of having contributions from both MSPs and DM annihilations,
the MSPs plus the $b\bar{b}$ model always provides better fits than the MSPs
	plus the $\tau^{+}\tau^{-}$ model. 
	Comparing the former to the pure DM $b\bar{b}$ model, 
	the MSP component is found to be marginally needed.
\end{enumerate}

\begin{acknowledgements}
This research is supported by the Original Innovation Program of the Chinese 
Academy of Sciences (E085021002). Z.W. acknowledges the support by
the Basic Research Program of Yunnan Province
(No. 202201AS070005) and the National Natural Science Foundation of
China (12273033).
\end{acknowledgements}

\bibliographystyle{aasjournal}
\bibliography{ms}

\clearpage

\begin{table}
\begin{center}
\caption{Spectral fitting results for the GCE emission}
\label{tab:fitting}
\begin{tabular}{lcccccc}
\hline
	Spectral Model & $\chi^{2}_{min}$ & MSP number & DM $m_{\chi}$ & DM ${\langle \sigma v \rangle}$ & $p$-value \\
	&  &  & (GeV) & (10$^{-26}$ cm$^{3}$\,s$^{-1}$) \\ \hline
	MSP & 42.4 & 1661 (1415--1706)& \nodata & \nodata & 2$\times 10^{-4}$ \\
	DM  $b\bar{b}$ & 20.1 & \nodata & 110 & 3.6 & 0.4 \\
	DM  $\tau^{+}\tau^{-}$ & 37.6 & \nodata & 16 & 0.8 & 6$\times 10^{-3}$\\
	DM   $b\bar{b}$ $+$ $\tau^{+}\tau^{-}$ & 12.4 & \nodata & 68 & 2.8 & 0.8 \\
	MSP $+$ DM  $b\bar{b}$ & 14.0 & 1100 & 606 & 9.0 & 0.7 \\
	MSP $+$ DM  $\tau^{+}\tau^{-}$ & 17.1 & 1300 & 94 & 1.6 & 0.5 \\
\hline
\end{tabular}
\end{center}
\end{table}

\begin{table}
\centering
\tabletypesize{\footnotesize}
\tablecolumns{10}
\tablewidth{240pt}
\setlength{\tabcolsep}{2pt}
\caption{Stacking analysis results for 96 \gr\ MSPs}
\label{tab:stacking}
\begin{tabular}{lccccccc}
\hline
Band & $F^{up}/10^{-11}$ & TS & $F^{up}_{s}/10^{-8}$ \\
(GeV) & (GeV cm$^{-2}$ s$^{-1}$) &  & (GeV cm$^{-2}$ s$^{-1}$ sr$^{-1}$) \\
\hline
30.0--76.6 & 3.34 & 14 & 1.83\\
76.6--195.7 & 0.39 & 0 & 0.21\\
195.7--500.0 & 1.85 & 0 & 1.01\\
\hline
\end{tabular}
\vskip 1mm
\footnotesize{Note: $F^{up}$ is the energy flux ($E^{2} dN/dE$) upper limit 
	obtained from the stacking analysis. $F^{up}_{s}$ is the upper limit 
	scaled for 1661 MSPs in the GCE region.}
\end{table}

\end{document}